\documentclass[conference,letterpaper]{IEEEtran}

\usepackage[utf8]{inputenc} 
\usepackage[T1]{fontenc}
\usepackage{url}
\usepackage{ifthen}
\usepackage{cite}
\usepackage[cmex10]{amsmath} 

\usepackage{cite}
\ifCLASSINFOpdf
 
\else
   \usepackage[dvips]{graphicx}
 \fi
\usepackage[cmex10]{amsmath}
\usepackage{amssymb}
\usepackage{stfloats}
\usepackage{amsthm} 
\usepackage{wrapfig}
\usepackage{amssymb}
\usepackage{latexsym}
\usepackage{bm, bbm} 
\usepackage{epsfig}
\usepackage{graphicx}
\usepackage{epsfig}
\usepackage{multicol}
\usepackage{psfrag}
\usepackage{float}
\usepackage{cite}
\usepackage{subfigure}
\usepackage{color}

\newcommand{\Z}{\mathbb{Z}}

\newcommand{\matr}[1]{{#1}}

\newcommand{\tr}{\mathsf{T}}

\newcommand{\defeq}{\triangleq}
\newcommand{\cC}{\mathcal{C}}
\def\girth{{\rm girth}}

\newtheorem{lemma}{Lemma}
\newtheorem{theorem}[lemma]{Theorem}

\newtheorem{corollary}[lemma]{Corollary}

\theoremstyle{plain}

\newtheorem{PreDefinition}[lemma]{{\textbf{Definition}}}
    {\begin{PreDefinition}}{\hfill$\square$\end{PreDefinition}}

\theoremstyle{plain}

\newtheorem{PreRemark}[lemma]{{\textbf{Remark}}}
  \newenvironment{remark}%
    {\begin{PreRemark}\upshape}{\hfill$\square$\end{PreRemark}}

\newtheorem{PreExample}[lemma]{{\textbf{Example}}}
  \newenvironment{example}%
    {\begin{PreExample}\upshape}{\hfill$\square$\end{PreExample}}

\long\def\symbolfootnote[#1]#2{\begingroup%
\def\thefootnote{\fnsymbol{footnote}}\footnote[#1]{#2}\endgroup} 

\begin{document}
\title{
Necessary and Sufficient Girth Conditions for 
 Tanner Graphs of  Quasi-Cyclic LDPC Codes} 


\author{%
  \IEEEauthorblockN{Roxana~Smarandache}
  \IEEEauthorblockA{Departments
of Mathematics and Electrical Engineering\\ University of Notre Dame\\ Notre Dame, IN 46556, USA\\
                    Email: rsmarand@nd.edu}
  \and
  \IEEEauthorblockN{David G. M. Mitchell}
  \IEEEauthorblockA{Klipsch School of Electrical and Computer Engineering\\
New Mexico State University\\Las Cruces, NM, USA\\
                    Email: dgmm@nmsu.edu}}


\maketitle


\begin{abstract}
 This paper revisits  the connection between the girth of a protograph-based LDPC code given by a parity-check matrix and the properties of powers of the  product between the matrix  and its transpose  
 in order  to obtain the necessary and sufficient conditions for a code to have given girth between 6 and 12,  and to show how these conditions can be incorporated into simple algorithms to construct codes of that girth. To this end, we highlight the role that certain submatrices that appear in these products  have in the construction of codes of  desired girth.   
In particular, we show that imposing  girth conditions on a parity-check matrix is equivalent to imposing  conditions on a square  submatrix  obtained from it and we show how this equivalence is particularly strong for a protograph based parity-check matrix of  variable node degree 2,  where the cycles in its  Tanner graph  correspond one-to-one to the cycles in the Tanner graph of a square submatrix obtained by adding the permutation matrices (or products of these) in the composition of the parity-check matrix. 
We end the paper with exemplary constructions of codes with various girths and computer simulations.     Although, we mostly assume the case  of 
fully connected protographs of  variable node degree 2 and 3,  the results can be used for any parity-check matrix/protograph-based Tanner graph.  
   \end{abstract}\vspace{-1mm}

%

\section{Introduction}\label{sec:intro}

%
%
%
%


Low-density parity-check (LDPC) codes, in particular quasi-cyclic LDPC (QC-LDPC) codes, are now found in many industry standards. One of the main advantages of QC-LDPC codes is that they can be described simply, and as such are attractive for implementation purposes 
since they can be encoded with low complexity using simple feedback shift-registers \cite{lcz+06} and their structure leads to efficiencies in decoder design \cite{wc07}. The performance of an LDPC  code with parity-check matrix $H$ depends on cycles in the associated Tanner graph, since cycles in the
graph cause correlation during iterations of belief propagation decoding \cite{ru01}. Moreover, these cycles form substructures found in the undesirable trapping and absorbing sets that create the  error floor. Cycles have also been shown to decrease the upper bound on the minimum distance (see, e.g., \cite{sv12}). 
Therefore, codes with large girth are desirable for good performance (large minimum distance and low error floor).  Significant effort has been made to design QC-LDPC code matrices with large minimum distance and girth, see \cite{klf01,tss+04,kncs07,phns13,kb13,msc14} and references therein.

In this paper, we will use some previous results by McGowan and Williamson \cite{mw03} and the terminology introduced in  Wu et al. \cite{wyz08}  that  elegantly relate  the girth of $H$ with the girth of $\matr{B}_n(H)\defeq \left(\matr{H}\matr{H}^\tr\right)^{\lfloor{n/2}\rfloor}\matr{H}^{(n\mod 2)}, n\geq 1$. We take  this connection in a new direction. Our purpose is to showcase  certain submatrices of $HH^\tr$  of importance when looking for  cycles in the Tanner graph of $H$ and thus to  highlight the role that these matrices have in the construction of codes of  desired girth.  
  In particular, we show that the cycles in the Tanner graph  of a $2N\times n_vN$ parity-check matrix $H$ based on the $(2, n_v)$-regular fully connected (all-one) protograph, with lifting factor $N$, correspond one-to-one to the cycles in the Tanner graph of a $N\times N$ matrix, that we call  $C_{12}$,  obtained from $H$. Similarly, we show that  imposing  girth conditions on a $3N\times n_vN$ parity-check matrix is equivalent to imposing girth conditions on a $3N\times 3N$ submatrix of  $HH^\tr$, which we call $C_H$.  Although we mostly assume the case  of  an  $(n_c, n_v)$-regular fully connected protograph, for  $n_c=2, 3$,   the results can be used to analyze the girth of the Tanner graph of any parity-check matrix.
  
   We use the results to construct codes of girth 6, 8, 10, and 12. We also show that, by following a two-step lifting procedure called pre-lifting \cite{msc14}, girth 12 codes can be pre-lifted in a deterministic way in order to obtain a girth 14 code and to increase the minimum distance. We conclude the paper with computer simulations of some of these codes, confirming the expected robust error control performance. We emphasize that we do not visit other constructions found in the literature because what we present is a unifying  framework, in particular providing necessary and sufficient conditions for a given girth to be achieved, and thus all constructions must fit in this framework. The proposed algorithms to choose lifting exponents are extremely fast, in fact they can be evaluated by hand, and could display codes of a given girth for the smallest graph lifting factor $N$.

\section{Definitions, notations and background}\label{sec:background}

We use the following notation, 
for any positive integer $L$,
$[L]$ denotes the set $\{  1,2, \ldots, L \}$. 
As usual, an LDPC code $\cC$ 
is described as the null space of a 
parity-check matrix $\matr{H}$ 
 to which we associate a Tanner
graph~\cite{tan81} in the usual way. The girth $\girth(\matr{H})$ of a graph is the length of the shortest cycle in the graph. 

A protograph \cite{tho03,ddja09} is a small bipartite graph  represented by a parity-check or \emph{base} biadjacency matrix  $B$ with non-negative integer entries $b_{ij}$. The parity-check matrix $H$ of a protograph-based LDPC block code can be created by replacing each non-zero entry $b_{ij}$  by a sum of $ b_{ij}$ non-overlapping $N\times N$ permutation matrices and a zero entry by the $N\times N$ all-zero matrix. Graphically, this operation is equivalent to taking an $N$-fold graph cover, or ``lifting'', of the protograph. We denote the $N\times N$ circulant permutation matrix where the entries of the $N\times N$ identity matrix $I$ are shifted to the left by $r$ positions modulo $N$ as $x^r$.

We  use the elegant triangle operator 
introduced in \cite{wyz08} between any two non-negative integers $e,f$ to define $$d\defeq e\triangle f\defeq 
        \begin{cases} 
           1 
             & \text{if $e\geq 2,  f=0 $} \\
           0                                     
             & \text{otherwise}
         \end{cases},$$  
 and between  two $s\times t$ matrices $\matr{E}=(e_{ij})_{s\times t}$ and $\matr{F}=(f_{ij})_{s\times t}$ with non-negative integer entries  to  define the matrix $\matr{D} =(d_{ij})_{s\times t}\defeq \matr{E}\triangle \matr{F}$ entry-wise  as  $d_{ij}\defeq e_{ij}\triangle f_{ij}, \text{ for all } i\in [s], j\in [t].$

The following theorem found in \cite{mw03} and \cite{wyz08} describes an important connection between $\girth(H)$ and matrices $\matr{B}_n(H)\defeq \left(\matr{H}\matr{H}^\tr\right)^{\lfloor{n/2}\rfloor}\matr{H}^{(n\mod 2)}, n\geq 1 $ and offers some insight on the inner structure of the Tanner graph which  simplifies considerably the search for QC protograph-based codes with large girth and minimum distance. 

  
\begin{theorem}(\hspace{-0.01mm}\cite{mw03} and \cite{wyz08})\label{adjacent-cond} A Tanner graph of an LDPC code with parity-check matrix $\matr{H}$ has  $\girth(H)>2g$ if and only if 
 $\matr{B}_t(H)\triangle \matr{B}_{t-2}(H) =\matr{0}, t=2,3,\ldots, g.$ 
  \end{theorem}
 Lastly, we extend the  theorem  on cycles in all-one protographs from \cite{msc14} that gives the algebraic conditions imposed by a  cycle of length $2l$ in the Tanner graph of an all-one protograph to the more general case of any protograph.

    \begin{theorem}\label{th:cycle}
Let $\cC$ be a code 
   described by a protograph-based  parity-check matrix $\matr{H}$ 
   where each $(i,j)$ entry is the $N\times N$ zero matrix or a sum of non-overlapping $N\times N$ permutation matrices, denoted $\matr{P}_{ij}$.  
 Then, a $2l$-cycle in  the Tanner graph of  $\matr{H} $
exists if and only if there exists  a sequence of  permutation matrices  $ \matr{P}_{i_0j_0}, \matr{P}_{i_1j_0},  \matr{P}_{i_1j_1},  \matr{P}_{i_2j_1}, \ldots,   \matr{P}_{i_{l-1}j_{l-1}},  \matr{P}_{i_0j_{l-1}}$ (with no two equal adjacent permutations)   
%
 such that 
$\left(\matr{P}_{i_0j_0}\matr{P}_{i_1j_0}^\tr  \matr{P}_{i_1j_1} \matr{P}_{i_2j_1}^\tr\cdots  \matr{P}_{i_{l-1}j_{l-1}} \matr{P}_{i_0j_{l-1}}^\tr +I\right)\triangle 0\neq 0.$\footnote{ A $2l$-cycle in the Tanner graph of $\matr{H}$  is a lifted cycle of  a $2l$-cycle in the protograph, i.e., it visits  sequentially the groups of nodes of the same type in the lifted graph in the same order in which the cycle visits the nodes of the original protograph.} 
\end{theorem}

 \section{The case of a $(2, n_v)$-regular  protograph} \label{sec:2byn}

We start the results of this paper with the case of $2\times n_v$ base matrices because, although it has limited practical importance  in its own, it becomes essential when seen as part of a larger protograph since each $n_c\times n_v$ base matrix of girth $g$, with $n_c\geq 2$,  has $n_c\choose 2$  $2\times n_v$ base matrices that must have girth at least $g$.  

\begin{theorem} \label{2byn} Let $P_i$ denote permutation matrices, $ i \in [n_v],$ $ n_v\geq 3$. Let $\matr{H}=\begin{bmatrix} \matr{I} &\matr{I} &\cdots &\matr{I}\\ \matr{P_1}& \matr{P_2}&\ldots &\matr{P_{n_v}} 
\end{bmatrix}$ and $C_{21}=C_{12}^\tr\defeq \sum\limits_{i=1}^{n_v}\matr{P_i}.$ Then $\girth(\matr{H}) =2\ \girth (C_{21}) .$  
\end{theorem} 
\begin{IEEEproof} From Theorem  \ref{th:cycle}, the Tanner graph associated with $\matr{H} $
has a cycle of length $2l$
if and only if there exist indices  $i_1,i_2,\ldots,i_{l}\in \{1,\ldots,n_v\},$     such that $i_s\neq i_{s+1} $   and 
such that  
$ \matr{I}\matr{P}_{i_1}^\tr  \matr{P}_{i_2} \matr{I}^\tr \matr{I}\matr{P}_{i_3}^\tr \matr{P}_{i_4} \matr{I}^\tr \cdots  \matr{P}_{i_{l-1}}^\tr\matr{P}_{i_{l}} \matr{I}^\tr \triangle I \neq 0\Longleftrightarrow$ 
$\matr{P}_{i_1}^\tr  \matr{P}_{i_2} \matr{P}_{i_3}^\tr \matr{P}_{i_4} \cdots \matr{P}_{i_{l-1}}^\tr \matr{P}_{i_{l}}  \triangle I \neq 0.$
  Equivalently, there exist $m_1,m_2, \ldots, m_{l}$ such that 
$ P_{i_1}(m_{2}, m_{1})= P_{i_2} (m_{2}, m_{3})= \cdots=P_{i_{l}}(m_{l}, m_{1})=1,$ 
which is equivalent to having an $l$-cycle in $C_{21}.$  
\end{IEEEproof}


\begin{corollary} \label{cor-2byn} Let $P_i, Q_i$ be permutation matrices, $i\in  [n_v],$ $ n_v\geq 3$. Let  
$H= \begin{bmatrix} \matr{P_0} &\matr{P_1} &\cdots &\matr{P_{n_v}}\\ \matr{Q_0}& \matr{Q_1}&\cdots &\matr{Q_{n_v}} 
\end{bmatrix}$ and  $C_{21}= C_{12}^\tr\defeq \sum\limits_{i=1}^{n_v}P_i^\tr  Q_i.$
Then  $ \girth(H) =2\ \girth(C_{21}). $
\end{corollary} 
\begin{IEEEproof}  The graph of $H$ is equivalent to the graph of the matrix 
$ \begin{bmatrix} \matr{I} &\matr{I} &\cdots &\matr{I}\\ \matr{P_0^\tr Q_0}& \matr{P_1^\tr Q_1}&\ldots &\matr{P_n^\tr Q_n}\end{bmatrix}$ which, based on Theorem~\ref{2byn}  has twice the girth of $C_{21}$. 
\end{IEEEproof} 

\begin{example} \label{2by3} To insure that the matrix  $H=\begin{bmatrix} \matr{I} &\matr{I} &\matr{I}\\ \matr{I}& \matr{P_2}&\matr{P_3}\end{bmatrix}$ of size $2N\times 3N$  has girth $8$ we only need to choose matrices $P_2,P_3$ such that  the matrix $I+P_2+P_3$ has entries 0 or 1, while in order for $H$ to have girth 12, we need to choose $P_2,P_3$ such that the girth of $I+P_2+P_3$  has girth 6.  For example,  the $7\times 7 $ parity-check matrix of the cyclic projective code given by the parity-check polynomial matrix $1+x+x^3$ has girth 6 giving a $14\times 21$ matrix $H$ with girth 12.  
Since   the girth of $I+P_2+P_3$ cannot  exceed the upper bound $6$ if $P_2,P_3$ are circulant, we need to  take them non-circulant to obtain a larger girth.  The matrix  $H$ with 
$$P_2= \begin{bmatrix}x&0&0 \\0&x^{13}&0\\ 0&0&x^7\end{bmatrix} \text { and } P_3=  \begin{bmatrix}0&x&0 \\0&0&x^2\\ x&0&0\end{bmatrix}$$ has girth  8 for a circulant size $N=7$, girth 10  for  $N=11$, and girth 12  for $N=31$. 
%
Therefore, the modulo 31 polynomial matrix  (or the $6\cdot 31\times 9\cdot 31$ scalar parity-check matrix) constructed with these matrices $P_2, P_3$  
has girth 24. 
 \end{example}

  \section{The case of a $(3, n_v)$-regular  protograph} \label{sec:mbyn} 
  We now provide results for the case of a general $3\times n_v$  base matrix. These results will be used in Section~\ref{sec:g12} to form simple constructive algorithms. 

\begin{theorem}\label{3byn} Let $H$ 
define the  $3N\times n_vN$ parity-check matrix of a protograph-based LDPC code such that: $P_1=Q_1=I$ and 
\begin{align*}
 \matr{H}&=\begin{bmatrix} I&I&\ldots&I\\ P_1&\matr{P_{2}}  &\ldots&\matr{P_{n_v}}\\ Q_1&\matr{Q_{2}}&\ldots&\matr{Q_{n_v}}\end{bmatrix}, C_H\defeq  \begin{bmatrix} 0&C_{12} &C_{13}\\ C_{21}&0&C_{23}\\ C_{31}&C_{32} &0\end{bmatrix} 
\\ & \matr{C_{12}}=C_{21}^\tr \defeq \sum\limits_{j=1}^{n_v} P_{j}^\tr ,   \matr{C_{13}}=C_{31}^\tr \defeq \sum\limits_{j=1}^{n_v} Q_{j}^\tr,\\& \matr{C_{23}}=C_{32}^\tr \defeq  \sum\limits_{j=1}^{n_v} P_{j}Q_{j}^\tr. \end{align*} 
Then the following equivalences hold. \pagebreak
\begin{enumerate}
\item $ \girth(H)>4  \Leftrightarrow  \girth(C_{ij})>2 
  \Leftrightarrow C_H\triangle 0 =0;$ 
\item $ \girth(H)>6 
\Leftrightarrow 
C_H\triangle 0 =0 \text{ and }
  C_HH\triangle H=0;$

\item $\girth(H)>8 \Leftrightarrow \girth\left(C_H \right) =6\Leftrightarrow C_H^2\triangle I=0;$ 
\item $\girth(H) >10\Leftrightarrow \left\{ \begin{matrix}  \girth(C_H)=6\\C_H^2H\triangle (H+ C_HH)=0\end{matrix} \right.;$
\item $\girth(H) >12\Leftrightarrow \left\{ \begin{matrix}  \girth(C_H)=6\\C_H^3\triangle (I+C_H+C_H^2)=0\end{matrix} \right..$
\end{enumerate} 
\end{theorem}
\begin{IEEEproof} Note that  \begin{align*}&\matr{B}_2(H)=\matr{H}\matr{H}^\tr=n_vI+C_H, \quad 
  \matr{B}_3(H)=n_vH+C_HH,  \\
&\matr{B}_4(H)=(n_vI+C_H)^2,   \matr{B}_5(H)=(n_vI+C_H)^2H,  \\&
 \matr{B}_6(H)=(n_vI+C_H)^3, \text{ etc..}
 \end{align*} 
Then 1) $\matr{B}_2(H)\triangle I= 0 \Leftrightarrow C_H\triangle 0=0$; \\
2)  $\matr{B}_3(H)\triangle \matr{B}_1(H)= 0 \Leftrightarrow  C_HH\triangle H=0 $;\\
3)  $B_4(H)\triangle B_2(H) =0 \Leftrightarrow (n_vI+C_H)^2 \triangle  (n_vI+C_H)=0\\\Leftrightarrow C_H^2\triangle (I+C_H)=0. $ 
  A 2- or 4-cycle can happen in $C_H$  if and only if it happens in  one of the matrices $\begin{bmatrix} C_{12} &C_{13}\end{bmatrix}, $ $\begin{bmatrix} C_{21} &C_{23}\end{bmatrix},$ $ \begin{bmatrix} C_{31} &C_{32}\end{bmatrix}. $  Since $\girth\left(C_H \right) =6 $ is equivalent to $C_H^2\triangle I=0$ we obtain that the weaker condition $C_H^2\triangle (I+C_H) =0$ must hold. The conditions for $\girth(H) >10$ and $12$ follow the same approach and are omitted due to space constraints.
\end{IEEEproof} 
\begin{remark} 1)   A similar theorem can be stated for the case $n_c>3$,  however,  $\girth(C_H) >4$ is only a necessary but not  sufficient condition  for $H$ to have girth 10. 

2) Note that $n_c\geq 3$,  $\girth(C_H)\leq 6, $ while for  $n_c\geq 4$,  $\girth(C_H)\leq 4$,  no matter the matrix $H$.  
 \end{remark} 

 We exemplify these results on a $3\times 4$ base matrix lifted to a protograph-based parity-check matrix of girth 10 from \cite{msc14}. 
 
  \begin{example} 
 Let
$\matr{H}=\begin{bmatrix} \matr{I} &\matr{I} &\matr{I}&\matr{I}\\ \matr{I}& \matr{P_2}&\matr{P_3}&\matr{P_4}\\ \matr{I}& \matr{Q_2}&\matr{Q_3}&\matr{Q_4}
\end{bmatrix}\defeq$ 
 $$\left[\begin{array}{cc|cc|cc|cc} 
\matr{1} & 0& \matr{1} &0& \matr{1}&0& \matr{1}&0 \\ 
0& \matr{1} & 0& \matr{1} &0& \matr{1}&0& \matr{1}\\\hline
 \matr{1}& 0&\matr{x}&0&0&\matr{x^{10}}&0&\matr{x^{13}}\\ 
 0& \matr{1}&  0&x^5& \matr{x^{10}}&0& \matr{x^{13}}&0\\  \hline
 \matr{1}& 0&0&\matr{x^{7}}&\matr{x^{11}}&0&\matr{x^2}&0\\
 0& \matr{1}& x^7&0&0& \matr{x^{11}}&0& \matr{x^{4}}
 \end{array}\right].$$
 
\noindent The polynomial matrices $C_{ij}(x)$ and $C_H(x)$ associated with $H$, $C_{ij}$ and $C_H$ are as follows. 
 \begin{align*}& C_{21}(x)=C_{12}^\tr(x)= \begin{bmatrix} 1+x&x^{10}+x^{13} \\ x^{10}+x^{13} &1+x^5 \end{bmatrix}, \\
&C_{31}(x)=C_{13}^\tr(x)=\begin{bmatrix}1+ x^2+x^{11}& x^7\\ x^7&1 +x^{4}+x^{11}\end{bmatrix}, \\
&C_{23}(x)=C_{32}^\tr(x)=\begin{bmatrix} 1& x^{-6}+x^{-1}+x^{9}\\ x^{-1}+x^{-2}+ x^{11}&1\end{bmatrix}. 
\end{align*}
The girth of $C_H$ is 6. So the $3N\times 3N$ much denser $(8,8)$-regular matrix $C_H$ has girth 6 while, equivalently, the $(3,4)$-regular $H$ has girth 10, or larger. \end{example}
 
 \section{Constructing $(3, n_v)$-regular protograph-based QC-LDPC codes of given girth $g\leq 12$}\label{sec:g12}

In this section, we will show how the equivalent conditions from Section~\ref{sec:mbyn} can be used to construct  QC matrices
\begin{align} \label{H-circulant} \matr{H}(x)=\begin{bmatrix} 1&1&1&\ldots&1\\ x^{i_{1}}&x^{i_{2}} &x^{i_{3}}&\ldots&x^{i_{n_v}}\\ x^{j_{1}}&x^{j_{2}}&x^{j_{3}}&\ldots&x^{j_{n_v}}\end{bmatrix},{i_{1}}=j_1=0, \end{align}  such that they have girth $6\leq g\leq 12$. 
We work with the polynomial matrices $C_{ij}(x)$ and $C_H(x)$ associated with the QC-scalar matrices $C_{ij}$ and $C_H$, defined as 
\begin{align}\label{3-matricesC}&\left\{\begin{matrix}\matr{C_{12}}(x)=C_{21}^\tr(x) \defeq &\sum\limits_{l=1}^{n_v} x^{-i_{l}} \\  \matr{C_{13}}(x)=C_{31}^\tr(x) \defeq &\sum\limits_{l=1}^{n_v} x^{-j_{l}} \\  \matr{C_{23}}(x)=C_{32}^\tr(x) \defeq& \sum\limits_{l=1}^{n_v} x^{i_{l}-j_{l}} \end{matrix}\right. .
 \end{align}

\begin{theorem} \label{girths} Let $H(x)$ and $C_H(x)$ be defined as in \eqref{H-circulant} and  \eqref{3-matricesC}. Then 
\begin{enumerate} \item $\girth(H(x)) >4$ if and only if 
 each one of the sets  $\{i_1,   \ldots, i_{n_v}\},   \{j_1, \ldots, j_{n_v}\},$  $\{i_1-j_1, \ldots, i_{n_v}-j_{n_v}\}$ contains non-equal values. 
 \item $\girth(H(x)) >6$ if and only if, for any $l\in[n_v]$,   
each one of the three sets below contains non-equal values:
\begin{align*} 
&\{i_l-i_s \mid s\in[n_v], s\neq l \}\cup  \{j_l-j_t \mid t \in[n_v], t\neq l\}, \\
&\{i_s\mid s\in[n_v], s\neq l \}\cup \{i_t-j_t+j_l\mid t \in[n_v], t\neq l\},\\
&\{j_s\mid s\in[n_v], s\neq l \}\cup \{j_t-i_t+i_l  \mid t \in[n_v], t\neq l\}.\end{align*}\noindent  Equivalently, $\girth(H(x)) >6$ if and only if
$$
j_l\notin \{ i_l+ (j_t-i_s), i_s+ (j_t-i_t),  j_s +(i_t -j_t)\mid 1\leq s,t<l \}. $$ 

\item $\girth(H(x)) >8$ if and only if  
each two  of  the following sets of differences contain non-equal values: 
\begin{align*}&
\{i_u-i_v\mid u\neq v, u,v\in[n_v]\},\\& \{j_u-j_v\mid u\neq v, u,v\in[n_v]\}, \\&\{(i_u-j_u)-(i_v-j_v)\mid u\neq v, u,v\in[n_v]\}.
\end{align*}

\item $\girth(H(x)) >10$ if and only if,   for all $l\in [n_v]$, 
\begin{enumerate} 
\item each two  of  the four sets  contain non-equal values:
\begin{align*}&\{i_u-i_v\mid u\neq v, u,v\in[n_v], u\neq l\} , \\& \{j_u-j_v\mid u\neq v, u,v\in[n_v], u\neq l\}, \\&
 \{-j_u+ j_v-i_v+i_l\mid u\neq v, u,v\in[n_v], v\neq l\},\\&\{-i_u+i_v-j_v+j_l\mid u\neq v, u,v\in[n_v], v\neq l\}. \end{align*}

\item  each two  of  the four  sets contain non-equal values:
\begin{align*}& \{i_u-j_u+j_v \mid u\neq v, u,v\in[n_v], v\neq l\} , \\& \{i_u-i_v+i_l\mid u\neq v, u,v\in[n_v], v\neq l\}, \\&
 \{(i_u-j_u)-(i_v-j_v)+i_l\mid u\neq v, u,v\in[n_v], v\neq l\},\\& \{i_u-j_v+j_l\mid u\neq v, u,v\in[n_v], v\neq l\}.\end{align*}

\item each two  of  the four sets contain non-equal values:
\begin{align*}&\{j_u-i_u+i_v\mid u\neq v, u,v\in[n_v], v\neq l\} , \\& \{j_u-i_v+i_l\mid u\neq v, u,v\in[n_v], v\neq l\}, \\&
 \{j_u- j_v+j_l\mid u\neq v, u,v\in[n_v], v\neq l\},\\& \{j_u-i_u+i_v-j_v+j_l\mid u\neq v, u,v\in[n_v], v\neq l\}. \end{align*}
\end{enumerate} 
 \end{enumerate}
\end{theorem} 
\begin{IEEEproof}
1) In order to avoid 4-cycles, we impose $\matr{C_{ij}}(x)\triangle 0=0$, for all $1\leq i <j\leq 3$. Equivalently,  the claim  on the three sets above holds.  

2) In order to avoid 6-cycles, we impose, for all $l\in [n_v]$,   
$$ \left\{\begin{matrix} 
 C_{12}(x)x^{i_l}+ C_{13}(x)x^{j_l}\triangle 1 =0,\\
 C_{21}(x)+ C_{23}(x)x^{j_l}\triangle x^{i_l} =0, \\
C_{31}(x)+ C_{32}(x)x^{i_l}\triangle x^{j_l} =0,\end{matrix}\right. $$
which is equivalent to the conditions below,  from which the claim follows: for all $ l\in [n_v]$  and all $s,t\in [n_v]\setminus \{ l\} $, 
 $$\left\{\begin{matrix} 
 \sum\limits_{{s=1}\atop {s\neq l}}^{n_v} \left(x^{i_l-i_s}+x^{j_l-j_s} \right)\triangle 1 =0\\
 \sum\limits_{{s=1}\atop {s\neq l}}^{n_v} \left(x^{i_s}+x^{i_s-j_s+j_l} \right)\triangle x^{i_l} =0\\
 \sum\limits_{{s=1}\atop {s\neq l}}^{n_v} \left(x^{j_s}+x^{j_s-i_s+i_l} \right)\triangle x^{j_l} =0\end{matrix}\right.
 \Longleftrightarrow   
 \left\{\begin{matrix}  
 x^{i_l-i_s}\neq x^{j_l-j_t}  \\
 x^{i_s}\neq x^{i_t-j_t+j_l}\\
 x^{j_s}\neq x^{j_t-i_t+i_l}
 \end{matrix}\right..$$
Conditions 3 and 4 are obtained in similar fashion.
\end{IEEEproof} 

Since the condition sets from Theorem \ref{girths} have relatively few elements, they can be integrated into simple algorithms to generate the lifting exponents for the desired girth. For example, we present two exemplary recursive algorithms to  choose these exponents: Type A in which we alternately choose the exponents $i_1, j_1, i_2, j_2,\ldots, i_{n_v},j_{n_v}$; and Type B in which we first choose $i_1, i_2, \ldots, i_{n_v}$  and then choose $j_1, j_2,\ldots, j_{n_v}$. We state below the steps followed in our algorithms for girth 8 and for girth 10 codes.  
\underline{\em Algorithm Type A for girth 8}\\
  Step1:  Set  $i_1=0, j_1=0$. Set  $l=1$. \\
Step 2: Let $l:=l+1$.   Choose $i_{l} \notin \{  j_s +(i_t -j_t)\mid  1\leq s,t<l \}$ and then $j_{l} \notin \{i_l+ (j_t-i_s), i_s+ (j_t-i_t), \mid 1\leq s,t<l\}.$\\
Step 3:  If $l=n_v$ stop, otherwise, go to Step 2.

\underline{\em Algorithm Type B for girth 10}\\
Step1:  Set  $i_1=0$. Set  $l=1$.  \\
Step 2: Let $l:=l+1$. Let   $i_l \notin \{ i_u+ i_s-i_t \mid 0\leq  u, t, s  \leq l-1\}.  $ \\
%
 Step 3:  If $l=n_v$ stop, otherwise, go to Step 2.\\
%
Step 4:  Set  $j_1=0$. Set  $l=1$. \\
Step 5: Let $l:=l+1$. $j_l \notin \{j_u+j_s-j_t,   j_{u} + i_a-i_b, j_u+(j_s-i_s)+(j_t-i_t), i_l+(j_{u} -i_u)+( i_a-i_b) , i_l+(j_{u} -i_u) +(j_{v} -i_v)-(j_{s} -i_s), i_l+(j_{u} -i_u) + (j_s-j_t ) \mid  0\leq  a, b \leq n_v, u, s, t<l\} $. \\
Step 6:  If $l=n_v$ stop, otherwise, go to Step 5. 

 

\begin{example} \label{girth10-Example} We use the algorithm Type B for girth 10 to 
obtain the following  $(3, 8)$-regular protograph-based code $C$ of girth 10 with $H(x)$ from \eqref{H-circulant}.
We follow Steps 1-3 and choose $i_1=0$, $i_2=1$, $i_3-i_2\notin \{i_2-i_1 \}$, so we may choose $i_3=2i_2+1=3$. Similarly, $i_4 =2i_3+1=7$.
We can choose, e.g., $i_5=2i_4+1$, but in this case, this is not the smallest possible value $i_4$. So we instead choose $i_5=i_4+ 
 \min^* \left(\Z \setminus \{ i_4-i_3, i_4-i_2, i_4-i_1, i_3-i_2, i_3-i_1, \right. \\ \left. i_2-i_1\} \right) = 
7+ \min^* \left(\Z \setminus \{ 4, 6, 7, 2, 3, 1\}\right)=7+5=12.$\footnote{The $ \min^*$ operator returns the minimum positive value from a set.}
We continue in the same way, by choosing the minimum positive value not in the respective forbidden set,  to obtain:
%
$$ \begin{bmatrix} x^{i_1}&\cdots&x^{i_8}\end{bmatrix} =  \begin{bmatrix} 1& x&x^3&x^7&x^{12}&x^{20}&x^{30}&x^{44}\end{bmatrix}.$$ 
Therefore, $C_{12} =  1+ x+x^3+x^7+x^{12}+x^{20}+x^{30}+x^{44}$ has girth 6 over, e.g., $N=1+2\times 44=89$. ($N$ is chosen such that the negative differences are not equal to positive ones.)
 
 We now choose the row $ \begin{bmatrix} x^{j_1}&\cdots&x^{j_8} \end{bmatrix} $ 
%
following Steps 4 and 5 that will ensure the conditions of Theorem~\ref{girths} are satisfied.
The following matrix has girth 10 for $N=554$ (for example) was obtained with this algorithm
$$H=\begin{bmatrix} 1&1&1&1&1&1&1& 1\\1& x&x^3&x^7&x^{12}&x^{20}&x^{30}&x^{44}\\ 1& x^{66} &x^{461}& x^{106}& x^{144}& x^{194}& x^{274} &x^{385}\end{bmatrix}.$$    

\noindent Note that $C_{12}$, $C_{13}$, and $C_{23}$ all have girth 6, giving three $(2,8)$ codes of girth 12. 
 \end{example}
    
  \begin{remark} The smallest $N$ for which a code of girth 10 exists can also be computed from Theorem~\ref{girths} as:
$N_{\min}=\min^* \left(\Z \setminus \{i_a+i_b-i_c-i_d,  j_u-j_v +i_a-i_b, \right. \\ \left.j_u-j_v+ j_s-j_t, j_u-j_v +(j_s-i_s)-(j_t-i_t), \right. \\ \left.
(j_s-i_s)-(j_t-i_t)+ (j_u-i_u)-(j_v-i_v),\right. \\ \left. (j_s-i_s)-(j_t-i_t) +i_a-i_b,  \mid a,b,s,t, u,v \in[n_v]\}\right).$
Similarly, based on Theorem~\ref{girths},  we can obtain the minimum lifting factor $N_{\min}$  for each desired girth $g\leq 12$.
\end{remark}

%
%
%
%
%
 The following theorem allows a fast way to choose the lifting exponents by taking increasing values that are larger than the ones in the ``forbidden" sets. We provide a girth 10 statement, but similar rules can be obtained for other girths.
 \begin{lemma} \label{girth10-construction} Let $H(x)$ and $C_H(x)$ be defined as in \eqref{H-circulant} and  \eqref{3-matricesC}.  Let $i_l$ and $j_l$ be defined recursively as:
\begin{align*} &\left\{ \begin{matrix}  i_1= 0 \\ i_l=1+2i_{l-1}, l\geq 2\end{matrix} \right.\text{ and } \left\{ \begin{matrix}  j_1= 0, j_2=1+i_2+2i_{n_v}\\ j_l=1+ 2j_{l-1} +i_{l},  
l\geq 3. \end{matrix} \right.\end{align*}

\noindent Then the Tanner graph of the code with parity-check  matrix $H(x)$ has girth 10 for  some $N$ (which is not too large).   \end{lemma} 

\begin{example} We build a $(3, 7)$-regular matrix based on Lemma~\ref{girth10-construction} as
$$H(x)=\begin{bmatrix} 
1&1&1&1&1&1&1\\
1& x           &x^3       &x^7       &x^{15}  &x^{31}   &x^{63}\\ 
1& x^{128} &x^{260}& x^{528}& x^{1072}& x^{2176}& x^{4416} \end{bmatrix}=$$ $$ \begin{bmatrix} 
1&1&1&1&1&1&1\\
1& x           &x^3       &x^7       &x^{15}  &x^{31}   &x^{63}\\ 
1& x^{128} &x^{260}& x^{95}& x^{206}& x^{11}& x^{86} \end{bmatrix}$$    
 The first matrix  has girth 10 for $N=433$, or larger.  The second matrix obtained by reducing the exponents  modulo $N=433$ has the minimum value $N=347$ for which the girth is 10.  We write $260=-87$ and obtain   $$H(x)=\begin{bmatrix} 
1&1&1&1&1&1&1\\
1& x           &x^3       &x^7       &x^{15}  &x^{31}   &x^{63}\\ 
1& x^{128} &x^{-87}& x^{95}& x^{-141}& x^{11}& x^{86} \end{bmatrix}$$ which has the minimum value $N=327$  for which the girth is 10. We update  $-141= 186$ and $-87=240$ for $N=327$, and rewrite the matrix as  $$ \begin{bmatrix} 
1&1&1&1&1&1&1\\
1& x           &x^3       &x^7       &x^{15}  &x^{31}   &x^{63}\\ 
1& x^{128} &x^{240}& x^{95}& x^{186}& x^{11}& x^{86}  \end{bmatrix}. $$   
The minimum $N$ for which this matrix has girth 10 is now $N=278$. 
We note that  $N=278$ is not the minimum for which a code can be found (the minimum found with the algorithm is $N=219$), but it is easily  obtained by hand. 
\end{example}

The following is another example obtained using the algorithm Type B for girth 12 (omitted due to space constraints), where the values chosen are of some random non-forbidden values rather than the minimum value possible at each point. 
 \begin{example} The  matrix 
  $\matr{H}=\begin{bmatrix} I&I&I&I&I\\1& x&x^7&x^{18}&x^{44}\\ 1& x^{32} & x^{54}& x^{141}& x^{133}\end{bmatrix}\vspace{1mm}$ 
has girth 12 for $N=245$ (length $n=1225$), for example. 
 \end{example} 
  \section{Obtaining QC-LDPC codes with girth larger than 12 and/or increased minimum distance} \label{pre-lift}  
To achieve girth larger than 12 and/or a minimum distance larger than the known upper bound $(n_c+1)!$ \cite{md01}, we cannot  take $H$ in the form~\eqref{H-circulant},   so we need to consider permutation matrices $P_i$ and $Q_i$ such that some (at least) are not circulant.  In \cite{msc14}, we showed how to increase the minimum distance  by  composing them of a sub-array of circulant matrices by first choosing the pre-lifting protograph and then choosing the circulant matrices to be placed according to  this protograph. A similar method can be applied not only to increase the minimum distance, but to also to obtain codes with Tanner graph of girth 14 or larger.  We exemplify the process below.

\begin{example} Let $P_1=Q_1=I,$  and let  the indices in the matrices $P_2, \ldots, P_5$ be
$ \begin{matrix} [1,0, 0],  [3,9,17],  [39, 4, 11], [29, 59, 71]\end{matrix}$,  respectively,  according to the protograph $\begin{bmatrix} x & x& x^2& x^2\end{bmatrix}$,  this means that, e.g., $P_2$ has non-zero entries $x^1, x^0, x^0$ in the $3\times 3$ permutation matrix corresponding to $x$.  
The indices in the matrices $Q_2, \ldots, Q_5$ are $[118,32,209],$ $[136,479,A],$ $[290, B, 800],$  $[353,C ,-319],$ respectively,  according to the protograph    $\begin{bmatrix} x^2&e&x&1\end{bmatrix} $ where $e$  is a (non-circulant) permutation matrix with its non-zero  positions on $(1,3),(2,2),(3,1)$.  Substituting $A, B,$ and $C$  by 0 (masking) gives a girth 14 irregular code for $N=891$.  Choosing  any  of $A, B, C$ to be  non-zero restricts the girth to 12, because a $2\times 3$ all-one protograph is included. Substituting $A=1199,B=1239,C=-579$ gives a girth 12 code for which many  12-cycles  were eliminated by choosing the majority of the exponents to give an (irregular) $H$ of girth 14. 
Both codes are simulated for $N=891$ (or length $n=13,365$) in Section \ref{sec:sim}.
 \end{example}

The final example demonstrates a construction of a girth 14 regular code obtained from a 3-cover (prelifted all-ones $3\times 5$ base matrix) that meets the conditions. Here, we must ensure that the 3-cover does not have any  $2\times 3$ all-one submatrix.  
 \begin{example} Let $H$ such that 
the indices in the matrices $P_2, \ldots, P_5$ are 
$[1,0, 7],$ $ [3,5,11],$ $ [6, 23, 29],$ $[15, 19, 42]$, according to the protograph $\begin{bmatrix}  x&x&1&x^2\end{bmatrix}$, and 
$ Q_2, \ldots, Q_5$ are $ [25,64,9],$ $ [61,180,143],$ $ [94, 239, 256],$ $ [153,358,474] $  according to  $\begin{bmatrix}  1&x^2&x&x\end{bmatrix}$,  respectively, 
where the notation $[1,0,7]$, for example,  means  that $P_2$ has  $x^1, x^0, x^7$ in the nonzero entries of the $3\times 3$ permutation matrix $x$. This graph has girth 14 for, e.g., 
$N=903$ (or length $n=13,545$). 
 \end{example}
\section{Simulation results}\label{sec:sim}\vspace{-1mm}

\begin{figure}[h]
\begin{center}
\includegraphics[width=3.5in]{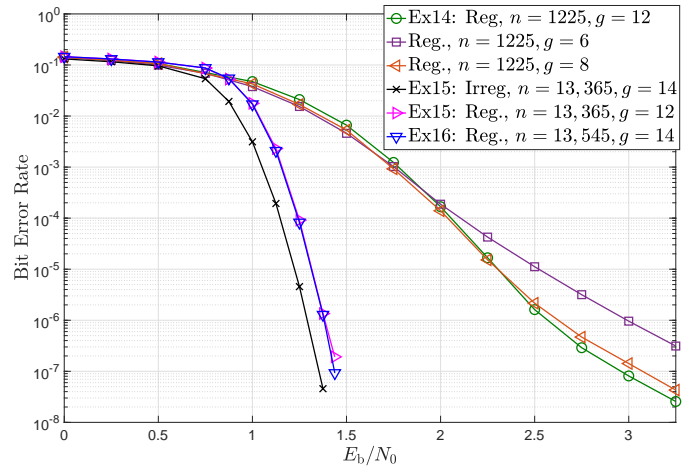}
\end{center}\vspace{-5mm}
\caption{Simulated decoding performance in terms of BER for the $R=2/5$ QC-LDPC codes from Examples 14-16.}\label{fig:comp}\vspace{-4mm}
\end{figure}

To verify the performance of the constructed codes, computer simulations were performed assuming binary phase shift keyed (BPSK) modulation and an additive white Gaussian noise (AWGN) channel. The sum-product message passing decoder was allowed a maximum of $100$ iterations and employed a syndrome-check based stopping rule. In Fig. \ref{fig:comp}, we plot the bit error rate (BER) for the $R\approx2/5$ QC-LDPC codes from Examples 14-16. Along with the performance of the $(3,5)$-regular QC-LDPC code with  girth 12  from Example 14, we show the  performance of constructed $(3,5)$-regular QC-LDPC codes of the same rate and length with girths 6 and 8. At lower SNRs, the higher girth codes perform slightly worse, but this ordering reverses in the error floor. With respect to the longer codes from Examples 15-16, we remark that the codes display no indication of an error-floor, at least down to a BER of $10^{-7}$. The regular codes from Examples 15 (reduced multiplicity of 12 cycles) and 16 (with girth 14) have similar performance in the simulated range, but we anticipate deviation at higher SNRs where the 12-cycles are involved in trapping sets.

\section{Concluding Remarks}\label{conclusion}\symbolfootnote[0]{This material is based upon work supported by the National Science Foundation under Grant Nos. OIA-1757207 and HRD-1914635. } In this paper we gave necessary and sufficient conditions for the Tanner graph of a protograph-based QC-LDPC code to have girth $6\leq g\leq 12$. We also showed how these girth conditions can be used to write fast  algorithms to construct such codes and exemplified them for  codes of girth 10. We also  showed that in order to exceed girth 12 a double graph-lifting procedure called pre-lifting can be employed, which  was demonstrated to construct QC-LDPC codes with girth 14.  

\bibliographystyle{IEEEtran}

\begin{thebibliography}{10}
\providecommand{\url}[1]{#1}
\csname url@rmstyle\endcsname
\providecommand{\newblock}{\relax}
\providecommand{\bibinfo}[2]{#2}
\providecommand\BIBentrySTDinterwordspacing{\spaceskip=0pt\relax}
\providecommand\BIBentryALTinterwordstretchfactor{4}
\providecommand\BIBentryALTinterwordspacing{\spaceskip=\fontdimen2\font plus
\BIBentryALTinterwordstretchfactor\fontdimen3\font minus
  \fontdimen4\font\relax}
\providecommand\BIBforeignlanguage[2]{{%
\expandafter\ifx\csname l@#1\endcsname\relax
\typeout{** WARNING: IEEEtran.bst: No hyphenation pattern has been}%
\typeout{** loaded for the language `#1'. Using the pattern for}%
\typeout{** the default language instead.}%
\else
\language=\csname l@#1\endcsname
\fi
#2}}

\bibitem{lcz+06}
Z.~Li, L.~Chen, L.~Zeng, S.~Lin, and W.~H. Fong, ``Efficient encoding of
  quasi-cyclic low-density parity-check codes,'' \emph{IEEE Transactions on
  Communications}, vol.~54, no.~1, pp. 71--81, Jan. 2006.

\bibitem{wc07}
Z.~Wang and Z.~Cui, ``A memory efficient partially parallel decoder
  architecture for quasi-cyclic ldpc codes,'' \emph{IEEE Transactions on Very
  Large Scale Integration (VLSI) Systems}, vol.~15, no.~4, pp. 483--488, Apr.
  2007.

\bibitem{ru01}
T.~J. Richardson and R.~L. Urbanke, ``Efficient encoding of low-density
  parity-check codes,'' \emph{IEEE Transactions on Information Theory},
  vol.~47, no.~2, pp. 638--656, Feb. 2001.

\bibitem{sv12}
R.~Smarandache and P.~O. Vontobel, ``Quasi-cyclic {LDPC} codes: Influence of
  proto- and {Tanner}-graph structure on minimum {Hamming} distance upper
  bounds,'' \emph{IEEE Transactions on Information Theory}, vol.~58, no.~2, pp.
  585--607, Feb. 2012.

\bibitem{klf01}
Y.~Kou, S.~Lin, and M.~P.~C. Fossorier, ``Low-density parity-check codes based
  on finite geometries: a rediscovery and new results,'' \emph{IEEE
  Transactions on Information Theory}, vol.~47, no.~7, pp. 2711--2736, Nov.
  2001.

\bibitem{tss+04}
R.~M. Tanner, D.~Sridhara, A.~Sridharan, T.~E. Fuja, and D.~J. {Costello, Jr.},
  ``{LDPC} block and convolutional codes based on circulant matrices,''
  \emph{IEEE Transactions on Information Theory}, vol.~50, no.~12, pp.
  2966--2984, Dec. 2004.

\bibitem{kncs07}
S.~Kim, {J.-S. No}, H.~Chung, and {D.-J. Shin}, ``Quasi-cyclic low-density
  parity-check codes with girth larger than 12,'' \emph{IEEE Transactions on
  Information Theory}, vol.~53, no.~8, pp. 2885--2891, Aug. 2007.

\bibitem{phns13}
H.~Park, S.~Hong, {J.-S. No}, and {D.-J. Shin}, ``Design of multiple-edge
  protographs for {QC LDPC} codes avoiding short inevitable cycles,''
  \emph{IEEE Transactions on Information Theory}, vol.~59, no.~7, pp.
  4598--4614, July 2013.

\bibitem{kb13}
M.~Karimi and A.~H. Banihashemi, ``On the girth of quasi-cyclic protograph
  {LDPC} codes,'' \emph{IEEE Transactions on Information Theory}, vol.~59, pp.
  4542--4552, 2013.

\bibitem{msc14}
D.~G.~M. Mitchell, R.~Smarandache, and {D. J. Costello, Jr.}, ``Quasi-cyclic
  {LDPC} codes based on pre-lifted protographs,'' \emph{IEEE Transactions on
  Information Theory}, vol.~60, no.~10, pp. 5856--5874, Oct. 2014.

\bibitem{mw03}
J.~McGowan and R.~Williamson, ``Loop removal from {LDPC} codes,'' in
  \emph{Proc. IEEE Information Theory Workshop}, Paris, France, 2003, pp.
  230--233.

\bibitem{wyz08}
X.~Wu, X.~You, and C.~Zhao, ``A necessary and sufficient condition for
  determining the girth of quasi-cyclic {LDPC} codes,'' \emph{IEEE Transactions
  on Communications}, vol.~56, pp. 854--857, 2008.

\bibitem{tan81}
R.~M. Tanner, ``A recursive approach to low complexity codes,'' \emph{IEEE
  Transactions on Information Theory}, vol.~27, no.~5, pp. 533--547, Sept.
  1981.

\bibitem{tho03}
J.~Thorpe, ``Low-density parity-check ({LDPC}) codes constructed from
  protographs,'' Jet Propulsion Laboratory, Pasadena, CA, INP Progress Report
  42-154, Aug. 2003.

\bibitem{ddja09}
D.~Divsalar, S.~Dolinar, C.~Jones, and K.~Andrews, ``Capacity-approaching
  protograph codes,'' \emph{IEEE Journal on Selected Areas in Communications},
  vol.~27, no.~6, pp. 876--888, Aug. 2009.

\bibitem{md01}
D.~J.~C. MacKay and M.~C. Davey, ``Evaluation of {Gallager} codes for short
  block length and high rate applications,'' in \emph{IMA Volumes in
  Mathematics and its Applications, Vol. 123: Codes, Systems, and Graphical
  Models}.\hskip 1em plus 0.5em minus 0.4em\relax Springer-Verlag, 2001, pp.
  113--130.

\end{thebibliography}

\end{document}